# Room temperature magnetodielectric studies on Mn doped LaGaO$_3$

Hari Mohan Rai[1], Ravikiran Late[1], Shailendra K. Saxena[1], Vikash Mishra[1], Rajesh Kumar[1] and P.R. Sagdeo[1*]

[1]Material Research Laboratory, Department of Physics and MSE; Indian Institute of Technology Indore, Indore (M.P.) – 452020, India

Archna Sagdeo[2]

[2]ISUD, Raja Ramanna Center for Advance Technology (RRCAT), Indore (M.P.) – 452013, India.

**Abstract:**

The polycrystalline samples of LaGa$_{1-x}$Mn$_x$O$_3$ (0$\leq$x$\leq$0.3) has been prepared by solid state reaction route. The phase purity of these samples has been confirmed by powder x-ray diffraction experiments carried out on BL-12 at Indus-2 synchrotron radiation source. The sample with x=0.2 shows significant change in the value of capacitance with the application of magnetic field. The observed results were understood by systematically analyzing magneto-capacitance (MC), magneto resistance (MR) and dielectric loss as a function of frequency. Our results and analysis suggests that the observed magneto dielectric (MD) coupling may be due to the MR effect of Maxwell-Wagner type and/or field induced dipolar relaxation. Further it is observed that the oxygen stoichiometry plays a very crucial role in observed MD coupling.
*Corresponding author: prs@iiti.ac.in

**Introduction:**

The magnetodielectric (MD) materials attracted considerable interest in recent past due to their rich physics and possible potential application in new generation of data storage systems. These materials belong to the category of highly correlated electron systems with strong electron-phonon, orbital, magnetic and ferroelectric coupling. In fact the normal state electronic, magnetic and ferroelectric properties of these materials continue to challenge our present understanding of solid state and condensed matter physics. In this context, huge efforts have been put by scientific community worldwide to understand the underlying physics of these materials. It is now well established that the strong coupling between spin and the orbital degrees of freedom plays a very crucial role in multiferroic coupling observed in these materials such as YMnO$_3$, BiFeO$_3$,

TbMnO$_3$, etc..[1] The discovery of ferroelectric polarization in BiFeO$_3$ and ferroelectric control of magnetic order in TbMnO$_3$ [2,3] has ignited intensive research activities in the field of magnetodielectric materials. [4] The co-occurrence of magnetic and ferroelectric orders makes such kind of materials very important. [5–7] The magnetoelectric (ME) effect has been witnessed in many magnetically induced ferroelectrics. [8,9] Likewise, magnetically induced reversal and rotation of electric polarization has been extensively explored in different material and structures. [3,10–12] But, the major limitation of such materials is that the observed MD coupling is weak and appears only at high magnetic field and below room temperature; [3,9–15] for example, 28 K in TbMnO$_3$,[3] 230 K in CuO [13] and 25 K in EuMnO$_3$,[14] this limits the practical application of these materials. However, room temperature ME effect has also been reported in Z-type hexaferrite-Sr$_3$Co$_2$Fe$_{24}$O$_{41}$; a polycrystalline ceramic sintered in oxygen. [16] These authors reported the ME effect at the operating frequency of 100 kHz with the application of magnetic field of ~0.8 T and attributed the origin of the observed MD effect to the cycloidal spiral spin arrangement in these materials. It should be noted that the MD effect has been observed mostly for commensurate and incommensurate spiral magnets; [17–19] hence crucial role of magnetic structure involving commensurate and incommensurate spiral magnets was proposed to be basic prerequisite for MD coupling. Further the crucial role of the symmetry of the crystal lattice with that of magnetic order and the role of the Dzyaloshinskii-Moriya interaction has been argued to be the important parameters for MD effect. [20,21] However G. Catalan [22] and Jang *et al* [23] provided the way to establish whether the observed MD effect is that of the intrinsic character of the sample or due to magnetoresistive effects or due to heterogeneous character of the sample. [24]

Keeping above in view, here, we present the room temperature MD studies on Mn doped LaGaO$_3$. It should be noted that pure LaGaO$_3$ is a well known dielectric material with very low value of dielectric loss. [25] Further, pure i.e. un-doped LaGaO$_3$ offers varieties of structural phase transformation as a function of temperature and pressure [25–27] and with Mn doping LaGaO$_3$ shows various magnetic-structures such as canted antiferromagnetic and ferromagnetic. [28] The small amount of Mn doping in LaGaO$_3$ is observed to induced ferromagnetic interactions in these samples with strong spin orbit correlations, [29] the oxygen stoichiometry can further modify the spin orbit interactions in these materials. [30] The Mn doped LaGaO$_3$ shows very high value of resistivity which is comparable to that of good dielectric systems. [31] Moreover, even though there exists significant literature on Mn doped LaGaO$_3$ [28–33] but to the best of our knowledge this material has not been studied for MD coupling. Keeping this in view, here, we investigate and report the effect of Mn doping at Ga site on the MD properties of LaGaO$_3$. Our results suggest that the Mn doped LaGaO$_{3+\delta}$ samples show significant MD response at room temperature.

**Experimental:**

Polycrystalline samples of LaGa$_{1-x}$Mn$_x$O$_3$ (x =0.05, 0.1, 0.15, 0.2 and 0.3) were prepared by conventional solid-state reaction route[28] with starting materials; La$_2$O$_3$ (99.99%), Ga$_2$O$_3$ (99.99%), and MnO$_2$ (99.98%). These starting materials were mixed in proper stoichiometric amount and grounded thoroughly with Propanol as a mixing medium. The resulting homogenous mixture was calcined in air ambient at 850 $^{\circ}$C, 1050 $^{\circ}$C, and 1200 $^{\circ}$C each time for 24 hours and final sintering was carried out at 1400 $^{\circ}$C in air for 24 hours with intermediate grindings. In order to examine the phase purity of the prepared samples the powder x-ray diffraction (XRD) experiments were carried out in angle dispersive mode using Huber 5020 diffractometer at BL-12 XRD beam line on Indus-2 synchrotron radiation source. The energy of the incidence x-ray beam was kept at 15 keV.[34] For dielectric measurements single phase powdered samples were pelletized at a high pressure of 15 ton to form 1 mm thick circular discs of 12 mm diameter and these pellets were sintered in air at 1400 $^{\circ}$C for 24 hours. The porosity of the prepared pellet samples were estimated and the same is found to be ~6%. These pellets were coated with silver paint and fired at 300 $^{\circ}$C for 30 minutes. Capacitance and magneto-capacitance of prepared pellets has been measured at room temperature using Wynne Kerr 65120B precision impedance analyzer with an oscillator voltage of $\pm$ 1 volt and in the absence and presence of magnetic field up to 0.4 Tesla. In order to avoid the contribution to MD phenomenon due to Hall Effect the direction of applied magnetic field was kept along the AC electric field used for dielectric measurements. We have ensured that there is no interference of applied magnetic field with the dielectric instruments. The room temperature M-H measurements were carried out on quantum design 14 Tesla PPMS-Vibrating Sample Magnetometer. The resistivity at 0 and 0.4 T magnetic field (MR) is measured using standard four probe method with Keithley 2182A nano voltmeter and 6221 source meter.

**Results and Discussions:**

Figure 1 shows the representative refined powder XRD pattern for polycrystalline sample of LaGa$_{0.8}$Mn$_{0.2}$O$_{3+\delta}$. This XRD pattern was refined using Fullprof package by considering orthorhombic structure with *Pnma* space group.[28] The values of the lattice parameters obtain from the refinements are a = 5.489(4) Å, b= 7.7686(4) Å and c = 5.5209(3) Å. The value for the goodness of fitting parameter was found to be 1.52. The absence of any unaccounted peak in the refined XRD patterns confirms the phase purity of the sample. This high purity sample was used for room temperature MD studies. It should be noted that the MD studies were carried out on all the prepared samples and this effect is observed only for samples with x = 0.2 and x = 0.3. For the sample with x=0.2, this MD effect is observed to be high and significant; hence in the next sections we have presented the result only for $LaGa_{0.8}Mn_{0.2}O_{3+\delta}$. Figure 2 shows the frequency dependence of $\varepsilon_r$ with and without magnetic fields i.e. 0 to 0.4 Tesla. The inset(s) shows magnified view for various frequency ranges. From this figure it is clear that there exists noticeable change in the value of $\varepsilon_r$ with the application of magnetic field. It is worth noticing that this significant change in value of $\varepsilon_r$ is observed at room temperature and for small value of applied magnetic field i.e. 0.4 Tesla. Further we have confirmed that this observed change is well above the sum of instrumental and statistical error bar even at the highest probing frequency (10 MHz). The room temperature magnetocapacitance (MC) response as a function of frequency is shown in figure 3; the same has been calculated by using following equation.

$$MC = \frac{\varepsilon_r(H) - \varepsilon_r(0)}{\varepsilon_r(0)} \times 100 \quad \text{-----} \quad (1)$$

Where, $\varepsilon_r(H)$ and $\varepsilon_r(0)$ represents the value of dielectric constant measured in the presence and absence of magnetic field respectively.

Broad peaks in the MC vs frequency plot are observed around 30 kHz for all applied magnetic fields as depicted in Figure 3. Further, this peak shifts systematically towards lower frequency side (see inset (i)) which points towards the presence of field induced dipolar relaxation.[24] It should be noted that all the measurements reported here, were repeated; the data and the observed difference in the magnetodielectric measurements are found to be consistent and well above the estimated error bar.

In order to understand the origin and to confirm, whether the observed magnetodielectric effect is intrinsic nature of the sample or it is arising due to indirect effect such as Maxwell-Wagner type (resistive origin), room temperature M-H measurements and four probe dc-MR measurements are carried out. The observed results are analyzed in the frame work proposed by Catalan [22] and Jang *et al.* [23]. Figure 4 shows the room temperature I-V plot for x=0.2 sample with and without application of magnetic field. From figure it is clear that the sample does not show any noticeable dc-MR at room temperature even at H=0.4 T. The inset of figure 4 shows room temperature M-H plot for the same sample, we could not observe any magnetization saturation even upto 14 Tesla, this suggests that the sample may be in paramagnetic state at room temperature which is consistent with the magnetization data reported by Blasco *et al.*[28] This paramagnetic nature of the sample discards the possibility of multiferroicity at room temperature. It will be really interesting to explore the multiferroic nature of these samples at various other temperatures. Wang et al.[35] and Ponomarev et al.[36] reported the magnetoelectric coupling in the paramagnetic state of a metal-organic framework and rare-earth molybdates respectively, these authors proposed that the observed magnetoelectric effect may be related to the magnetoelasticity/ magntostriction effect present in their samples, this need to be explored for the samples under present investigation.

In order to further confirm the origin of observed MD effect in our sample we have plotted the frequency dependence of dielectric loss (tanδ) with and without magnetic field as shown in the figure 5. The presence of two peaks in the tanδ, points towards the presence of two sub-types of dielectric relaxation, (by considering frequency range) i. e., space charge relaxation (first broad peak) and dipolar relaxation (second broad peak), [37, 38] in the sample under study. The value of the tanδ is ~1 for the frequencies in the kHz range, on the other hand, this value is exceptionally small (as shown in inset (iii)) ~0.08 for the frequencies above 1 MHz, which suggests that these samples may be useful for high frequency applications. With the application of the magnetic field the value of tanδ (leakage current) slightly increases throughout the probing frequency range. It should be noted that the dielectric loss (tanδ) is given by

$$\tan\delta = \frac{\sigma}{\varepsilon_r \omega} \quad \text{----- (2)}$$

From figure-2, figure-5 and equation-2 it is clear that with the application of magnetic field the conductivity of the sample (leakage current) increases throughout the frequency range. It should be noted at low frequency for polycrystalline samples the value of dielectric constant, and dielectric loss are governed by the contacts electrodes (interfacial space charge polarization), microstructure (grain boundary) and that of the intrinsic grain.[37, 38] The change in the value of tanδ and $\varepsilon_r$ is high for low frequency regime as compared to that of high frequency regime, suggesting that for the sample under investigation grain boundaries are more sensitive to the magnetic field than that of the grains[37, 38]. Thus the frequency dependent (*ac*) measurements as shown in figure 5 suggest that the observed change in the value of dielectric constant with the application of magnetic field may not be intrinsic property of the sample. G Catalan [22] has theoretically investigated the effect of negative MR on the intrinsic nature of magneto-dielectric materials with rigorous computational analysis. G Catalan [22] has shown that for a material with negative dc-MR, the peak in the tanδ as a function of frequency appears to be shifted towards higher frequency side with the application of magnetic field. It should be noted that for present case the value of dc-MR is very small (not measurable). Hence in order to search for the signature of MR as proposed by G Catalan[22] we have fitted the peak position in tanδ versus frequency data. The corresponding fitting is shown in the inset (iv) of figure-5 and the peak frequency positions, with and without magnetic field, are also tabulated in same figure. From the fittings it is clear that both the peaks (as marked by down arrows) shifts towards high frequency side with the application of magnetic field, which points towards the presence of negative MR in the sample [22,23] which is very common for manganites [39]. Thus it appears that the frequency dependent *ac* measurements (in present case) are more sensitive as compared to that of dc-MR measurements. The origin of negative value of MR may be understood on the basis of hopping of electron between Mn sites. [40] It should be noted that such hopping demands coexistence of $Mn^{3+}$ and $Mn^{4+}$ states in the same sample i.e. mixed valence states. The presence of Mn in +4 states in present sample is confirmed using iodometric titration experiments. Our titration experiments reveal that oxygen is present in excess in the sample as the value of δ is found to be +0.054. This suggests that ~54% of total Mn is present in $Mn^{4+}$ state. In order to further understand the effect of oxygen stoichiometry on the MD properties of this sample we have first properly removed the silver paint and annealed the same pellet in air ambient for 24 hours at 1400 °C with a slow heating and cooling rate of 1 °C/min. and repeated all the measurements as described above. We could not find the significant difference in x-ray diffraction pattern and the porosity of the sample, but the dielectric properties of the annealed sample are found to be significantly altered than that of the as prepared compound as depicted in figure 6. The annealing of sample results in an increase in the value of oxygen off-stoichiometry with δ = 0.06. Figure 6 (a), (b) and (c) show the dielectric constant, MC and tanδ respectively for annealed sample as a function of frequency with and without magnetic field.

Comparing the data shown in figure 6 for the annealed sample and that of as prepared sample (figures 2 to 5) it is clear that the oxygen stoichiometry plays very crucial role as it exceptionally modifies the value of dielectric constant, MC and tanδ.

For both the samples (as prepared and annealed) the peak in tanδ, with the application of magnetic field, shifts towards higher frequency side which points toward presence of negative MR[22]; on the contrary the MC peak frequency also shifts with application of magnetic field (field induced dipolar relaxation[24] ) which is in contrast with the results/arguments of resistive origin of magnetodielectric coupling; this requires further investigations. The experiments on high quality stoichiometric samples or experiments on high quality single crystal samples (to minimize the grain boundary contributions) may be useful in this regard. Further it will be interesting to study the ferroelectric and magnetic properties of the stoichiometric/single crystal samples.

**Conclusions:**

We report room temperature magnetodielectric studies on polycrystalline samples of $LaGa_{0.8}Mn_{0.2}O_{3+\delta}$ using very low magnetic fields i.e. from 0.1 T to 0.4 T. Our results and analysis suggests that the oxygen stoichiometry plays very crucial role in dielectric and magneto-dielectric properties of these samples. The presence of negative MR and field induced dipolar relaxation in the present samples are evident from the peak shift observed in the tan$\delta$ and that of MC respectively with the application of magnetic.

It should be noted that the appearance of magnetoresistance in the present sample is observed only through the AC measurements where as the dc-MR was almost zero (not measurable). Thus in order to have a more clear view about the resistive origin of MD coupling it is recommended that the analysis of ac MR along with the dc MR measurements is necessary. Further we propose that the value of negative MR may be minimized by making the samples with $\delta$ close to zero. Finally we introduce $LaGa_{0.8}Mn_{0.2}O_{3+\delta}$ as a new MD compound at room temperature.

**Acknowledgements:**
Authors would like to acknowledge, Prof. Gustau Catalan for valuable communications. Authors sincerely thank Prof. P. Mathur, Director IIT Indore for his encouragement. SIC IIT Indore is acknowledged for providing some of the experimental facilities. We sincerely thank Raja Ramanna Center for Advance Technology (RRCAT) Indore for providing synchrotron radiation facilities and in particular Dr. G.S. Lodha for his encouragement. Authors are grateful to Dr. Alok Banarjee of UGC-DAE CSR and Prof. A.K. Nigam of TIFR Mumbai for magnetization measurements. We sincerely acknowledge Dr. N. P. Lalla and Dr. Kiran Kumar of UGC-DAE CSR for the valuable discussion and suggestions. Authors sincerely thank Dr. A. K. Sinha and Mr. Manvendra N. Singh for their help during x-ray diffraction measurements. Smt. Aarti Deshpande from RRCAT is acknowledged for providing important references. Mr. Aga Shahee is acknowledged for extending help in determining the oxygen stoichiometry through titration method. One of the authors (HMR) acknowledges the Ministry of Human Resource

Development (MHRD), government of India for providing financial support as Teaching Assistantship. PRS would like to thank DAE-BRNS for funding the Impedance Analyzer used for the present measurements.

**Figure Captions:**

**Figure-1:** Rietveld refined powder x-ray diffraction pattern for 20 % Mn doped $LaGaO_3$ collected on Indus-2 synchrotron radiation source. The most intense peaks are indexed considering orthorhombic structure with *Pnma* space group. Absence of any unaccounted peak in the pattern confirms the phase purity of the sample.

**Figure-2:** Room temperature dielectric constant (for the sample with δ= +0.054) as a function of frequency for different applied magnetic fields. The effect of magnetic field on dielectric constant is evident from the magnified views shown in the insets for respective frequency ranges.

**Figure-3:** Room temperature Magnetocapacitance (for the sample with δ= +0.054) as a function of probing frequency for different magnetic fields. Inset (i) shows the variation of MD peak frequency as a function of magnetic field. The inset (ii) presents the MD data at higher frequencies with error bar for H=0.4T.

**Figure-4:** Current-Voltage (*I-V*) graph plotted by measuring the voltage for different currents at room temperature in the absence (H = 0 T ■) and presence (H = 0.5 T ○) of magnetic field using four-probe method. The inset shows room temperature M-H curve for the sample up to 14 T.

**Figure-5:** Room temperature Dielectric loss (for the sample with δ= +0.054) as a function of frequency for different applied magnetic fields. The insets (i), (ii) and (iii) shows the magnified views of corresponding frequency ranges. The inset (iv) shows the fitting of two broad peaks observed in tanδ.

**Figure-6:** (a) Dielectric constant (b) Magnetocapacitance and (c) Dielectric loss (tanδ) as a function of frequency (for the sample with δ= +0.06) measured at room temperature in the absence (H = 0 T ▲) and presence (H = 0.4 T ○) of magnetic field. The insets of (a) represent the effect of magnetic field on the value of dielectric constant for different frequency ranges. The insets of (b) and (c) show the magnified view of high frequency MC peak and tanδ peak respectively.

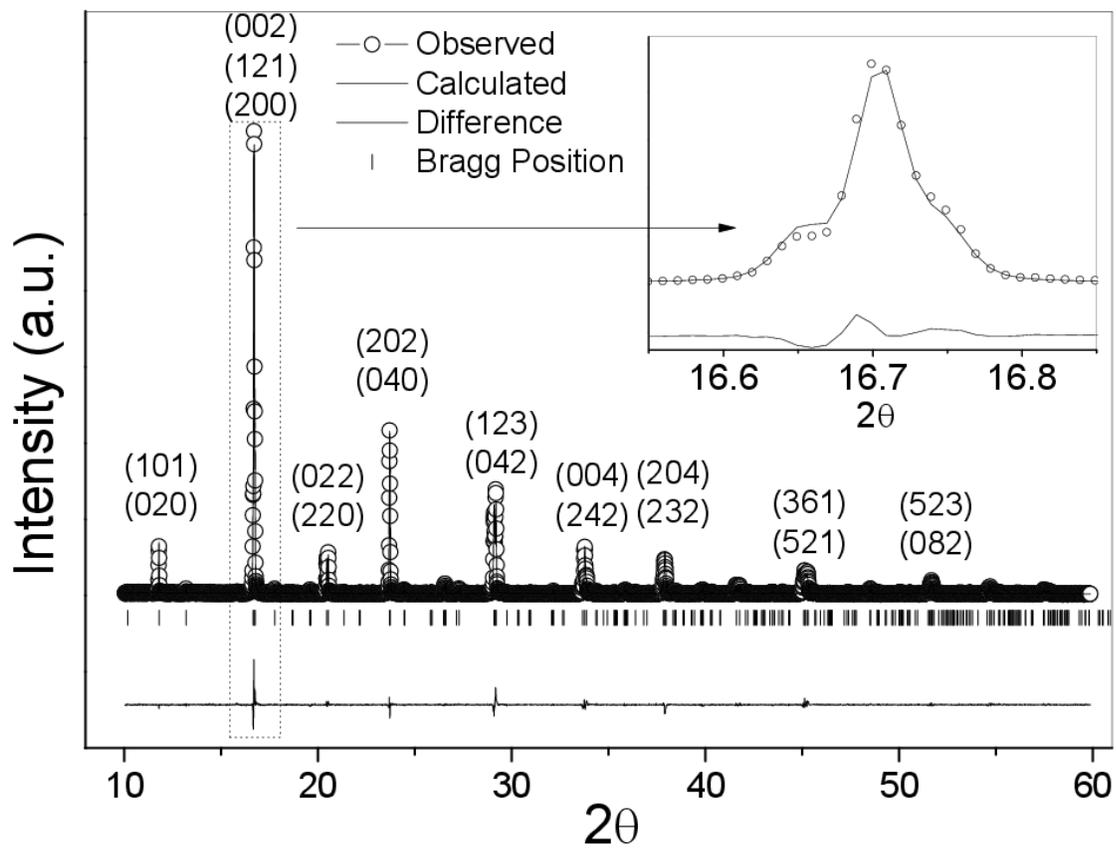

**Fig; 1    Rai.** *et al.*

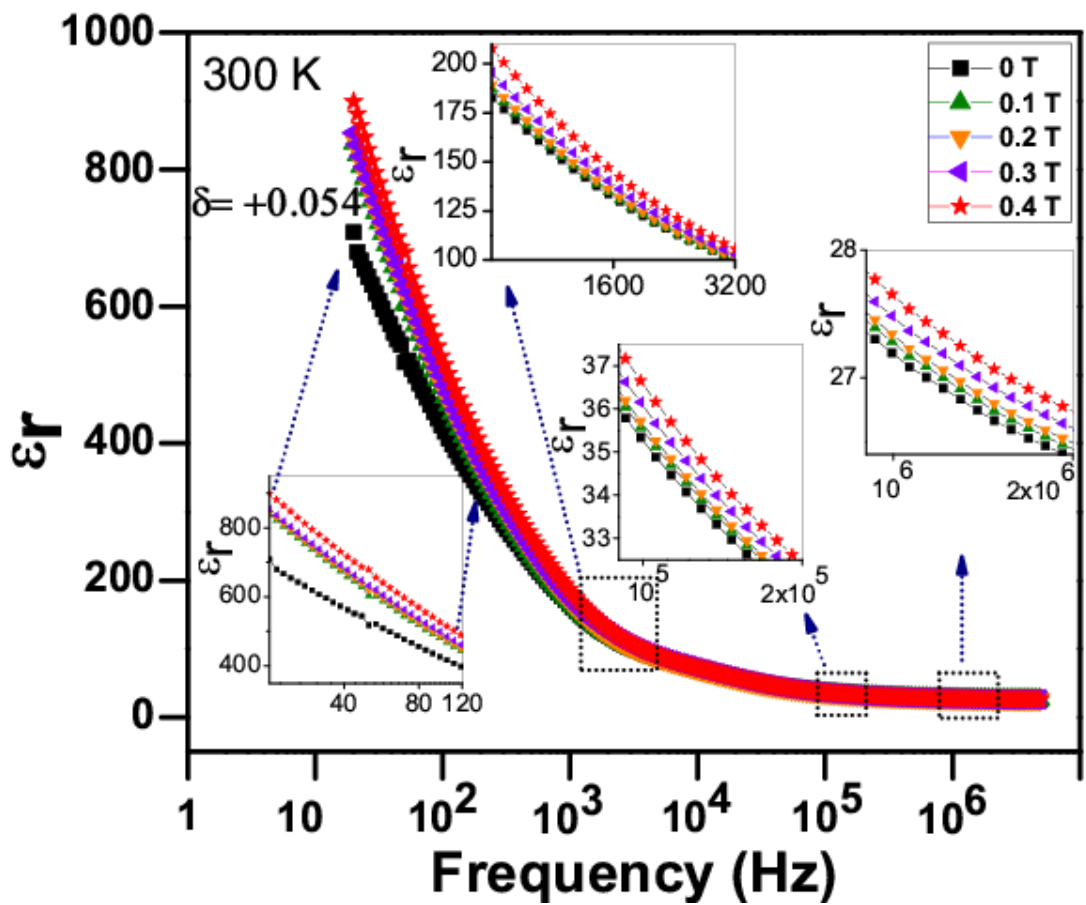

**Fig; 2    Rai. *et al.***

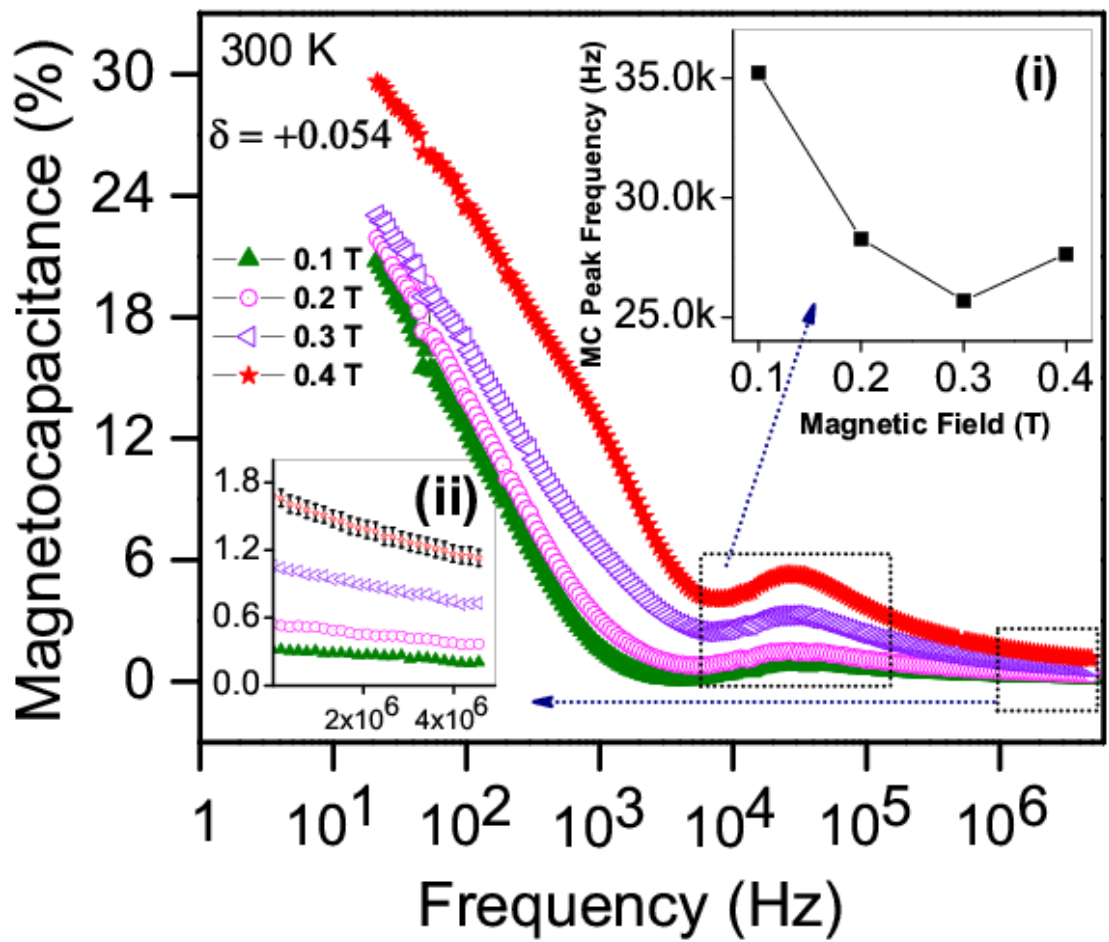

**Fig; 3    Rai.** *et al.*

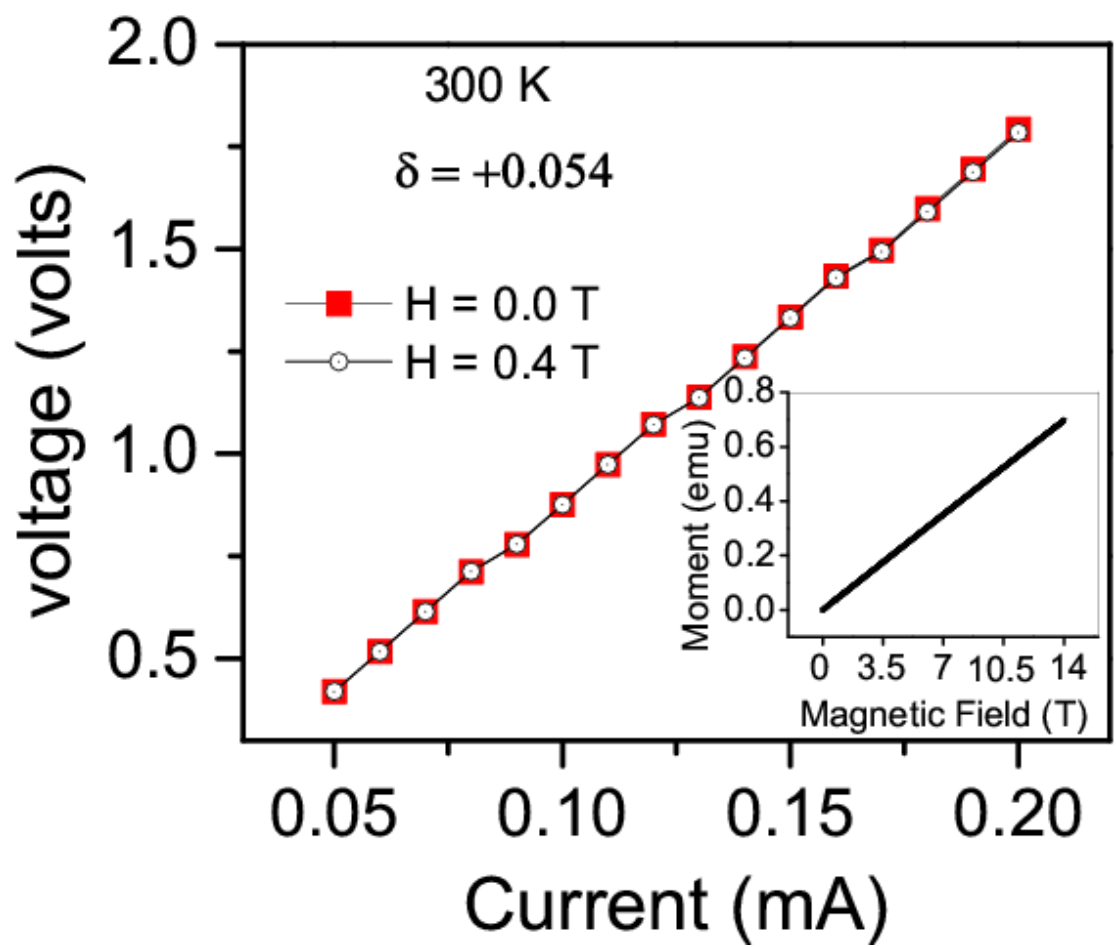

Fig; 4    Rai. *et al.*

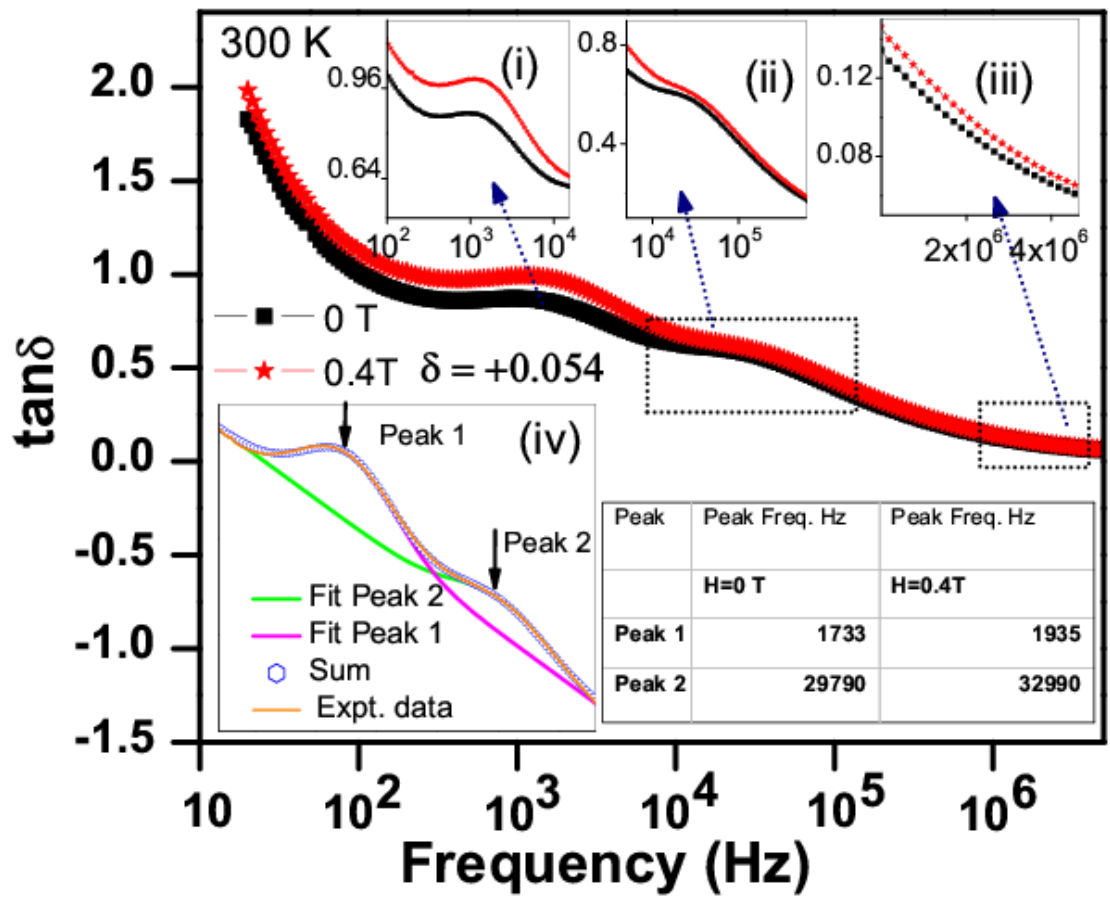

Fig; 5    Rai. *et al.*

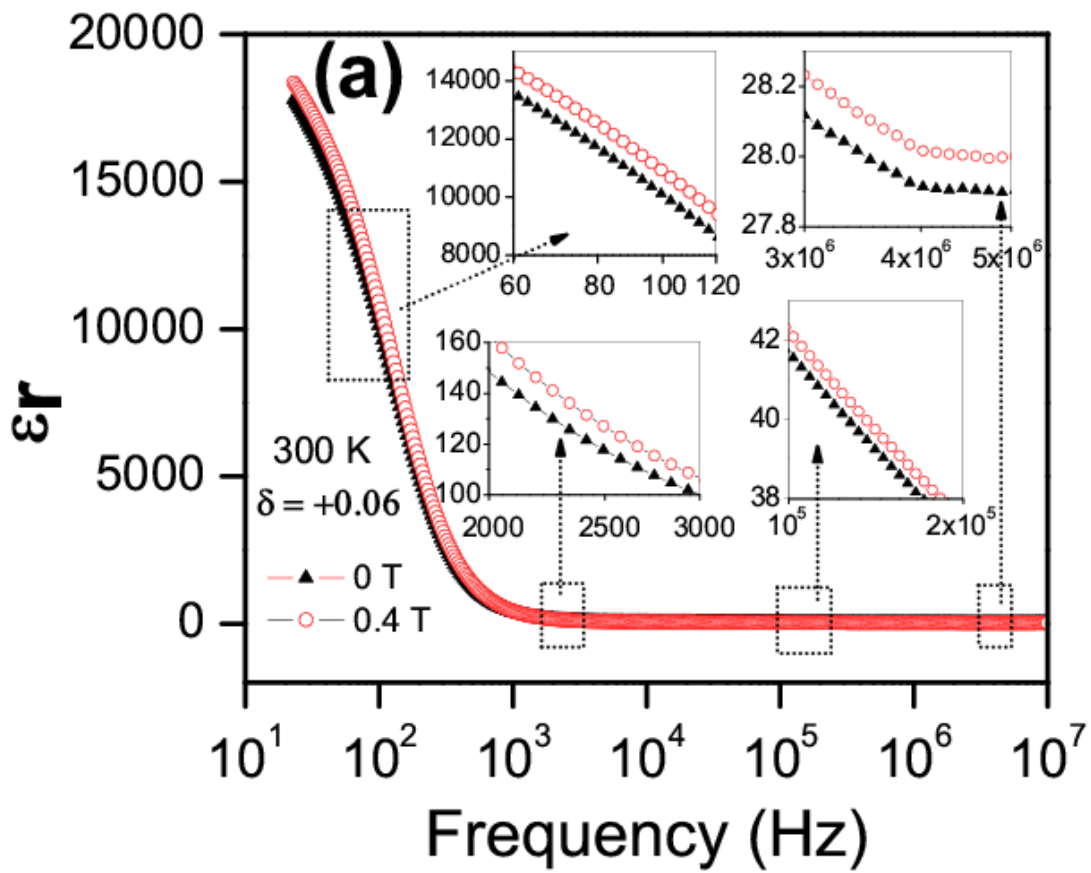

Fig; 6(a)    Rai. *et al.*

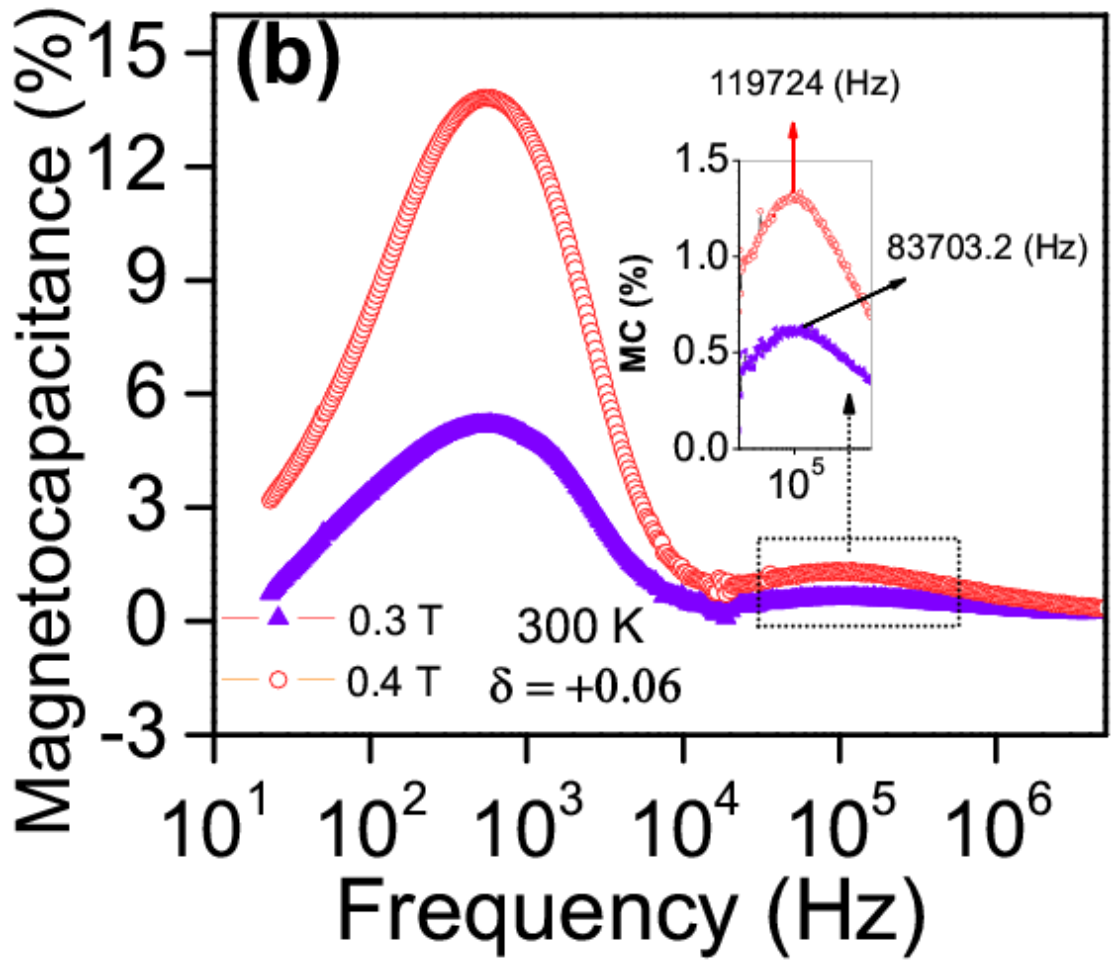

**Fig; 6(b)        Rai.** *et al.*

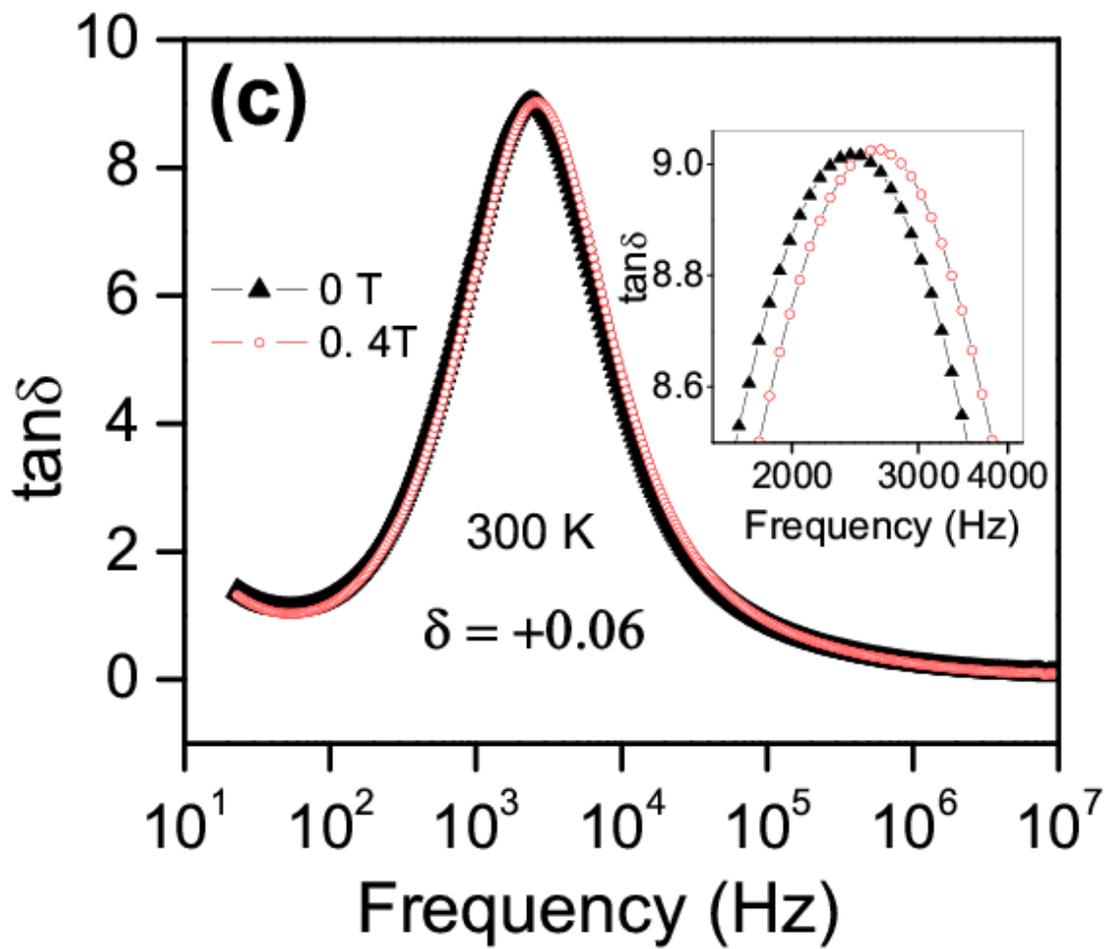

Fig; 6(c)   Rai. *et al.*